\documentclass[journal=jacsat,manuscript=article]{achemso}

\usepackage{chemformula} 
\usepackage[T1]{fontenc} 
\usepackage{multirow}
\usepackage{subfig} 
\usepackage{hyperref}
\usepackage{soul}

\newcommand{\abs}[1]{\left|#1\right|} 

\usepackage{color}

\author{Yanyu Duan}
\affiliation{Thrust of Advanced Materials, and Guangzhou Municipal Key Laboratory of Materials Informatics, The Hong Kong University of Science and Technology (Guangzhou), Guangzhou 511453, China.}

\author{Zecheng Gan} 
\email{zechenggan@hkust-gz.edu.cn; zechenggan@ust.hk}
\affiliation{Thrust of Advanced Materials, and Guangzhou Municipal Key Laboratory of Materials Informatics, The Hong Kong University of Science and Technology (Guangzhou), Guangzhou 511453, China.}
\alsoaffiliation{Department of Mathematics, The Hong Kong University of Science and Technology, Hong Kong SAR, China}

\title[An \textsf{achemso} demo]
  {Quantitative Theory for Critical Conditions of Like-Charge Attraction Between Polarizable Spheres}

\abbreviations{IR,NMR,UV}

\begin{document}


\begin{abstract}
Despite extensive experimental and theoretical efforts, a concise quantitative theory to predict the occurrence of like-charge attraction (LCA) between polarizable spheres remains elusive. 
In this work, we first derive a novel three-point image formula, based on a key observation that
connects the classical Neumann's image principle with the incomplete beta function.
This approach naturally yields simple yet precise critical conditions for LCA, with a relative
discrepancy of less than $1\%$ compared to numerical simulations, validated across diverse
parameter settings.
The obtained critical conditions may provide physical insights into various processes potentially involving LCA, such as self-assembly, crystallization, and phase separation across different length scales.
Additionally, the new image formula is directly applicable to enhance the efficiency of polarizable force field calculations involving polarizable spheres.

\end{abstract}


\textit{Introduction.} Electrostatic effect plays a significant role in nature across various length scales~\cite{levin2002electrostatic,grosberg2002colloquium,Elemessina2009electrostatics,Eleohshima2024fundamentals}.Examples include interactions between biomolecules\cite{DNAhe2023sequence,DNAkornyshev2007structure}, dust charging and transport~\cite{Dustwang2016dust,Dustbaptiste2021influence}, aerosol growth in Titan's atmosphere~\cite{Aerosollindgren2017effect}, and the self-assembly of charged colloidal particles~\cite{SAleunissen2005ionic,SAbarros2014dielectric}. 
One of the fundamental principles that describes the electrostatic interaction between electrically charged particles is Coulomb's law~\cite{coulomb1785premier,jackson2021classical}. 
According to this classical law, like-charged particles repel, oppositely-charged particles attract. 
However, for charged particles at short distances, Coulomb's law may not accurately describe the interaction because polarization effects between the particles can become significant.

One counterintuitive phenomenon resulting from polarization effects at short distances is like-charge attraction (LCA)~\cite{ZSbesley2023recent,Naturelee2023direct}. In recent years, LCA has been reported in the electrostatic interactions between charged conducting spheres~\cite{LCAlekner2012electrostatics}, polarizable spheres~\cite{ICMxu2013electrostatic,MSFqin2016theory}, rough dielectric particles~\cite{BEMgorman2024electrostatic}, and particles in a uniform external field~\cite{LCAli2024like}.
Conversely, LCA can also occur for like-charged particles in electrolyte solutions~\cite{LCAlarsen1997like,LCAallahyarov1998attraction,LCAlevin1999nature,LCAlinse1999electrostatic,LCAmoreira2001binding,LCAbaumgartl2006like,LCAnagornyak2009mechanism,LCAzhao2016like,JCTCjiang2021revisiting,Bbudkov2022modified,LCAwills2024anti,Bbudkov2024surface,Naturewang2024charge}, where both polarization and electrostatic correlations among mobile ions become crucial~\cite{zhang2005long}.
Studies indicate that LCA significantly influences various physical chemistry processes, including cation mobility in ionic liquids~\cite{khudozhitkov2023like}, polyelectrolyte self-assembly~\cite{zaki2016unexpected}, protein adsorption at the silica–aqueous interface~\cite{mcumber2015electrostatic}, and ion-pairing in water~\cite{vazdar2012like}.

Considerable effort has been devoted to explaining this counterintuitive phenomenon driven by polarization effects, utilizing various numerical methods and theoretical models~\cite{ZSbesley2023recent}. 
Developed numerical methods include
the finite element method~\cite{FEMfeng2000electrostatic}, boundary element method~\cite{BEMruan2022surface,BEMgorman2024electrostatic,hassan2022manipulating}
, multilevel method~\cite{MCsaunders2021new}, method of moments~\cite{MEbichoutskaia2010electrostatic,MEderbenev2016electrostatic,MEsiryk2022arbitrary}, image charge method~\cite{ICMwang2013effects,ICMxu2013electrostatic,LCAli2024like}, and hybrid methods~\cite{HMgan2016hybrid,HMgan2019efficient}. Additionally, several theoretical models have been proposed, including the bi-spherical coordinate transformation approach~\cite{BCkhachatourian2014electrostatic,BClian2018polarization}, polarizable ions model~\cite{PIchan2020theory}, and multiple-scattering formalism~\cite{MSFfreed2014perturbative,MSFqin2016image,MSFqin2016theory,MSFqin2019charge}. 

Despite the extensive numerical and theoretical investigations mentioned above, a concise and quantitative theory to predict the occurrence of LCA remains elusive.
In this Letter, we propose such a theoretical model for systems consisting of two like-charged polarizable spheres.
We address the challenge of close interaction between two polarizable spheres by deriving a new three-point image formula, based on which a simple yet accurate theory for the critical conditions of LCA is obtained.
The remainder of this paper is structured as follows. First, the classic Neumann's image principle for the polarizable sphere model is revisited. Then we derive the new three-point image formula and present a general theory for the critical conditions of LCA.
Finally, we discuss the critical conditions of LCA for both equal-sized and unequal-sized spheres, considering asymmetries in permittivity and carrying charges within each sphere.

\begin{figure}[htbp]
    \centering
    \includegraphics[scale=0.64]{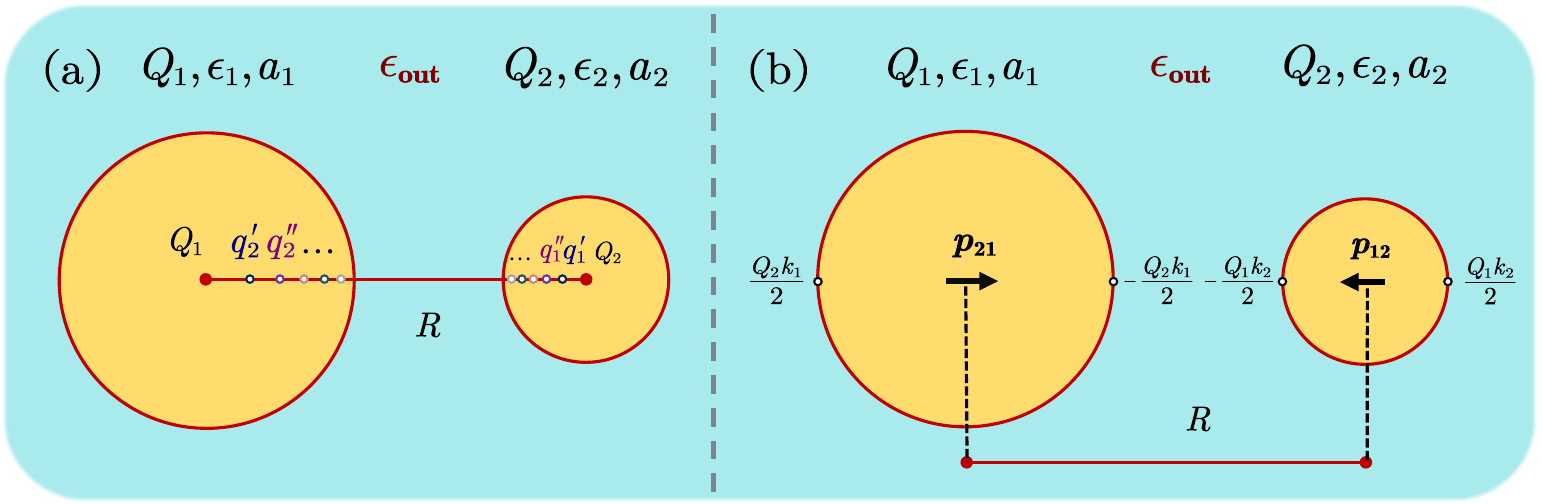}
    \caption{Schematics of the system setup: (a). the reflected Neumann's image charges; and (b). the three-point image formula (developed in this work), for two polarizable spheres immersed in a medium. In both panels, colored tiny hollow circles denote image charges, while arrows indicate dipole moments.}
    \label{schematic}
\end{figure}

\textit{The polarizable sphere model and Neumann's image principle.} Consider two polarizable spheres immersed in a homogeneous dielectric medium with permittivity $\epsilon_{\mathrm{out}}$, each sphere having a radius $a_i$, relative dielectric constant $\epsilon_i$ ($\epsilon_i>\epsilon_{\mathrm{out}}$), and central charge $Q_i$ ($i\in\{1,2\}$), separated by a center-to-center distance $R$, as illustrated in Fig.~\ref{schematic}. Such polarizable sphere model has been extensively studied in the modeling and simulation of various systems across different length scales, including polarizable ions~\cite{levin2009polarizable}, charged colloids~\cite{SAbarros2014dielectric}, and biomolecules~\cite{lotan2006analytical}.

A classical approach for evaluating the polarization potential and field is the image charge method. 
Let us start by considering a single, charge-neutral polarizable sphere with radius $a$ and relative dielectric constant $\epsilon_{\mathrm{in}}$, suspended in a medium characterized by relative permittivity $\epsilon_\text{out}$.
The so-called \emph{Neumann's image principle}~\cite{neumann1883hydrodynamische,lindell1993application} has been derived to account for the polarization potential induced by an external point charge $Q$.  
Suppose the point charge $Q$ is located at a distance $R$ from the center of the sphere ($R>a$), then the polarization potential outside the sphere can be expressed as the sum of the Coulombic potential generated by a Kelvin image charge $Q_\text{K}=-\frac{Qka}{R}$, positioned at the Kelvin inversion point $r_\text{K}=\frac{a^2}{R}$ inside the sphere, and a Neumann line image density $q_\text{line}(r)=\frac{Qkg}{a}\left(\frac{r_\text{K}}{r}\right)^{1-g}$ distributed from the sphere center to the Kelvin point (i.e., $r\in[0, r_\text{K}]$), where we define $k=\frac{\epsilon_\text{in}-\epsilon_\text{out}}{\epsilon_\text{in}+\epsilon_\text{out}}$ and $g=\frac{\epsilon_\text{out}}{\epsilon_\text{in}+\epsilon_\text{out}}$. 
It can be verified that $Q_\text{K} + \int_{0}^{r_\text{K}} q_\text{line}(r)dr = 0$, thereby ensuring that the total charge neutrality condition within the polarizable sphere is maintained.
Consequently, the polarization energy $E_{\mathrm{pol}}$ for the single-sphere system can be represented as 
\begin{equation}
    E_{\mathrm{pol}} =\frac{1}{2}\frac{Q}{4\pi\varepsilon_0\epsilon_\text{out}} \left( \frac{Q_\text{K}}{R-r_\text{K}} + \int_{0}^{r_\text{K}} \frac{q_\text{line}(r)}{R-r} \, dr\right)\;,
    \label{ICM1}
\end{equation}
where $\varepsilon_0$ is the absolute vacuum permittivity value, and the $1/2$ prefactor is due to the fictitious nature of image charges. 
In numerical simulations, tailored quadrature schemes have been developed to approximate the integral of $q_\text{line}$ in Eq.~\eqref{ICM1} by a few discrete point charges.
Such \emph{multiple-image} approximation was first proposed by W. Cai et al.~\cite{cai2007extending}, and subsequently applied to Monte Carlo (MC) and Molecular Dynamics (MD) simulations of polarizable sphere systems~\cite{gan2011multiple,gan2015comparison}.

Next, consider the two-sphere system, the polarization potential can be constructed through an iterative process of image-charge reflections, as depicted in Fig.~\ref{schematic}~(a). 
The first-level images inside each sphere are induced by the central charge of the other sphere. Subsequently, each first-level image induces multiple second-level images according to the Neumann's image principle, as described above. 
This reflection process generates an infinite series of images charges, all aligned along the center-to-center axis of the spheres (also see Fig.~\ref{schematic}~(a)). 
Numerically, since both $\abs{k}$ and $a/R$ are less than 1, the image strength decays exponentially as the reflection level increases, and this infinite recursion can be truncated once a specified tolerance is achieved.
The image-charge reflection approach has been applied to polarizable force field calculations in simulations of charged colloidal suspensions~\cite{gan2015comparison}, as well as to quantitative investigations of polarization-induced LCA phenomena in systems of two dielectric spheres carrying discrete surface charges~\cite{ICMxu2013electrostatic}, or in a uniform external field~\cite{LCAli2024like}.

\textit{The three-point image formula.} While the image-charge reflection approach facilitates quantitative calculations of polarization contributions, the complexity of its infinite series representation poses a significant challenge in developing a concise theory to predict the occurrence of LCA between two like-charged polarizable spheres.
Here, we derive a novel three-point image formula to replace the infinitely reflected images, allowing the development of a concise and quantitative theory to determine the critical conditions of LCA.

First, by substituting the definitions of $Q_\text{K}$,  $r_\text{K}$ and $q_\text{line}$ back into Eq.~\eqref{ICM1} and rearranging it, we obtain
\begin{equation}
     E_{\mathrm{pol}} = \frac{1}{2}\frac{Q}{4\pi\varepsilon_0\epsilon_\text{out}} \left[ -\frac{Qka}{R^2-a^2} + \frac{Qkg}{a} \left(\frac{a^2}{R}\right)^{1-g} \int_{0}^{\frac{a^2}{R}} \frac{r^{g-1}}{R-r} \, dr\right]\;.
    \label{ICM3}
\end{equation}
Next, we introduce two new dimensionless variables $t=\frac{a}{R}$ and $u=\frac{r}{R}$ (clearly, $0<u<t<1$ is always satisfied). By substituting these into Eq.~\eqref{ICM3}, we obtain
 \begin{equation}
     E_{\mathrm{pol}} =  \frac{1}{2} \frac{Q}{4\pi\varepsilon_0\epsilon_\text{out}}\left[ -\frac{ Qka}{R^2-a^2} + \frac{Qkg}{a} t^{2-2g} \int_{0}^{t^2} \frac{u^{g-1}}{1-u} \, du\right]\;.
    \label{ICM5}
\end{equation}
To further simplify Eq.~\eqref{ICM5}, we recall the \emph{incomplete beta function} $B_\alpha(p, q)$, customarily defined for any $0\leq \alpha \leq 1$, $p>1$ (if $\alpha=1$, also $q>0$) as~\cite{arfken2011mathematical}
\begin{equation}
     B_\alpha(p, q) =\int_0^\alpha u^{p-1}(1-u)^{q-1}du\;.
    \label{betadef}
\end{equation}
By setting $\alpha = t^2$, $p=g$ and $q=0$ in Eq.~\eqref{betadef}, we can express the integral term in Eq.~\ref{ICM5} using the incomplete beta function as follows:
\begin{equation}
    E_{\mathrm{pol}} = \frac{1}{2} \frac{Q}{4\pi\varepsilon_0\epsilon_\text{out}} \left[ -\frac{Qka}{R^2-a^2} + \frac{Qka}{R^2} \frac{g}{t^{2g}} B_{t^2}(g,0)\right]\;.
    \label{ICM6}
\end{equation}
To further simplify the complexity of Eq.~\eqref{ICM6}, we introduce the following series expansion for $B_{t^2}(g,0)$~\cite{pearson1968tables},
\begin{equation}
	\begin{split}
		B_{t^2}(g,0)&=t^{2g} \sum_{n=0}^{\infty}\frac{t^{2n}}{n+g} \\
		&=t^{2g}\left(\frac{1}{g}+\frac{t^{2}}{1+g}+\frac{t^{4}}{2+g}+...\right)\;.
	\end{split}
	\label{beta}
\end{equation}
It is important to note that the series expansion in Eq.~\eqref{beta} converges for $0 < t^2 <1$ and $g>0$, these conditions are consistently met under our system settings.
Now by substituting Eq.~\eqref{beta} into Eq.~\eqref{ICM6}, we obtain
\begin{equation}
    E_{\mathrm{pol}} = \frac{1}{2} \frac{Q}{4\pi\varepsilon_0\epsilon_\text{out}} \left[ -\frac{Qka}{R^2-a^2} + \frac{Qkag}{R^2}  \left(\frac{1}{g}+\frac{t^{2}}{1+g}+\frac{t^{4}}{2+g}+...\right)\right]\;.
    \label{ICM6p5}
\end{equation}
Finally, we decompose the first term in Eq.~\eqref{ICM6p5} using partial fraction, yielding
\begin{equation}
    E_{\mathrm{pol}} = \frac{1}{2} \frac{Q}{4\pi\varepsilon_0\epsilon_\text{out}} \left[ \left(\frac{Qk}{2}\right)\frac{1}{R+a} + \left(-\frac{Qk}{2}\right)\frac{1}{R-a} + \frac{Qkag}{R^2}  \left(\frac{1}{g}+\frac{t^{2}}{1+g}+\frac{t^{4}}{2+g}+...\right)\right]\;.
    \label{ICM6p6}
\end{equation}
Interestingly, by defining an image dipole moment $\mathbf p$, oriented from the sphere center to the point source (due to axial symmetry), and strength 
\begin{equation}
  p=|\mathbf{p}|=Qkag\left(\frac{1}{g}+\frac{t^{2}}{1+g}+\frac{t^{4}}{2+g}+...\right)\;,
    \label{dipole}
\end{equation}
we obtain a novel three-point image formula, expressed as:
\begin{equation}
    E_{\mathrm{pol}} = \frac{1}{2} \frac{Q}{4\pi\varepsilon_0\epsilon_\text{out}} \left[ \left(\frac{Qk}{2}\right)\frac{1}{R+a} + \left(-\frac{Qk}{2}\right)\frac{1}{R-a} + \frac{p}{R^2}\right]\;.
    \label{ICMfinal}
\end{equation}
Clearly, Eq.~\eqref{ICMfinal} indicates that, the polarization energy for a point charge outside a single polarizable sphere can be understood as contributed by three images (as also shown in Fig.~\ref{schematic}~(b)). 
Unlike Neumann's image principle, the new formula comprises two image charges and one image dipole: a pair of charges with strength $\pm Qk/2$ positioned on opposite sides of the sphere, and an image dipole situated at the sphere center.
Similar to Neumann's image principle, two key physical properties are satisfied: 1). all images are aligned along the same line to maintain axial symmetry; and 2). the total charge-neutrality condition is preserved.

Now consider the two-sphere system, the polarization energy can still be determined by the aforementioned infinite image reflection process. The advantage of Eq.~\eqref{ICMfinal} over Neumann's image principle is clear: with recursive reflections, all image charges and dipoles consistently remain positioned at the same three points. 
Consequently, contributions need only be accumulated at these three points throughout the reflection process, significantly reducing the complexity in theory and computation.

\begin{figure}[htbp]
	\centering
	\includegraphics[scale=1.3]{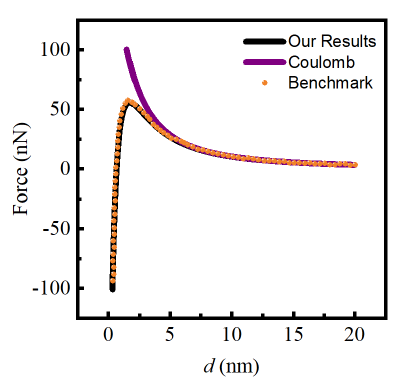}
	\caption{The interaction force between two polarizable spheres in vacuum as a function of sphere-sphere separation $d=R-a_1-a_2$. Both spheres have a radius of $a_1=a_2=1.25 nm$, carry central charges of $Q_1 = -1e$ and $Q_2 = -7e$, and dielectric constants $\epsilon_1 = \epsilon_2 = 20$. Black curve: this work; orange dots: benchmark results from Ref~\cite{MEderbenev2016electrostatic}; purple curve: the bare Coulomb interaction between the two spheres.}
	\label{validate}
\end{figure}

To validate the new image formula, we cross-compare our results with benchmark values from a previous study~\cite{MEderbenev2016electrostatic}, where a harmonic expansion representation for the interaction between two polarizable spheres was developed. 
In Fig.~\ref{validate}, we plot the electrostatic force between two charged polarizable spheres in vacuum. Both spheres have a radius of $a_1=a_2=1.25$nm, carry central charges of $Q_1 = -1e$ and $Q_2 = -7e$, and dielectric constants $\epsilon_1 = \epsilon_2 = 20$. 
As can be seen in Fig.~\ref{validate}, LCA occurs at short separation distances, resulting from the combined effects of polarization and charge-asymmetry in the two-sphere system.
Clearly, our theory demonstrates excellent agreement with previous methods. 
And given the simplicity of the three-point image formula, it holds the potential to enhance the efficiency of polarizable force field calculations in many relevant applications. Validation data cross-compare with Z. Xu's work~\cite{ICMxu2013electrostatic} is provided in the \emph{Supporting Information} (SI). 

\textit{Critical conditions for LCA: general theory.}  
To construct a concise and general theory to predict the occurrence of LCA between two polarizable spheres, we start with make an approximation to the dipole moment in Eq.~\eqref{dipole}. By only keeping the leading order term, i.e., $p \approx Qka$, the three-point image formula Eq.~\eqref{ICMfinal} becomes,
\begin{equation}
     E_{\mathrm{pol}} \approx \frac{1}{2} \frac{Q}{4\pi\varepsilon_0 \epsilon_\text{out}}   \left[\left(\frac{Qk}{2}\right)\frac{1}{R+a} + \left(-\frac{Qk}{2}\right)\frac{1}{R-a} + \frac{Qka}{R^2}\right]\;.
    \label{ICM8}
\end{equation}
Physically, it is understood that LCA occurs for strongly polarizable spheres ($\epsilon_{1,2} \gg \epsilon_{\mathrm{out}}$), which means that $g=\frac{\epsilon_\text{out}}{\epsilon_{1,2}+\epsilon_\text{out}}\ll 1$, and also $0<t<1$, thus one can conclude that the leading order term in Eq.~\eqref{dipole} is dominant over the next-to-leading order term, i.e., $1/g \gg t^2/(1+g)$. 
Cross-comparing with benchmark results will be discussed next, justifying the validity of the approximation used here.

Now consider the two-sphere system, by applying Eq.~\eqref{ICM8} up to the first-level image reflection (as shown in Fig.~\ref{schematic}~(b)), the total electrostatic interaction energy $E_{\mathrm{ele}}$ can be expressed as
\begin{equation}
		\begin{split}
          E_{\mathrm{ele}}= \frac{1}{4\pi\varepsilon_{0}\epsilon_\text{out}}\left\{   \frac{Q_{1}Q_{2}} {R}+\frac{1}{2}Q_{1}\left[\left(k_2 Q_{1}a_{2}\right) \frac{1}{R^2}+\left(\frac{Q_{1}k_{2}}{2}\right) \frac{1}{R+a_{2}}+\left(-\frac{Q_{1}k_{2}}{2}\right) \frac{1}{R-a_{2}}\right] \right.  \\ \left.
			+\frac{1}{2}Q_{2}\left[\left(k_1 Q_{2}a_{1}\right) \frac{1}{R^2}+\left(\frac{Q_{2}k_{1}}{2}\right) \frac{1}{R+a_{1}}+\left(-\frac{Q_{2}k_{1}}{2}\right) \frac{1}{R-a_{1}} \right] \right\}
		\end{split}\;.
		\label{energy}
\end{equation}
Then the electrostatic force $F$ exerted on the right sphere ($F>0$ replusive; $F<0$ attractive) follows from $F=-\partial E_{\mathrm{ele}}/\partial R$: 
\begin{equation}
    \begin{split}
		F=\frac{1}{4\pi\varepsilon_{0}\epsilon_\text{out}} \left\{
		\frac{Q_{1}Q_{2}} {R^{2}}+Q_{1}\left[\left(k_2 Q_{1}a_{2}\right) \frac{1}{R^{3}}+\left(\frac{Q_{1}k_{2}}{4}\right) \frac{1}{(R+a_{2})^{2}}+\left(-\frac{Q_{1}k_{2}}{4}\right) \frac{1}{(R-a_{2})^{2}}\right]  \right.  \\ \left.
			+Q_{2}\left[\left(k_1 Q_{2}a_{1}\right) \frac{1}{R^{3}}+\left(\frac{Q_{2}k_{1}}{4}\right) \frac{1}{(R+a_{1})^{2}}+\left(-\frac{Q_{2}k_{1}}{4}\right) \frac{1}{(R-a_{1})^{2}} \right] \right\}
		\end{split} \;.
		\label{force}
\end{equation}
While certainly known to experts, we believe that it has never been stated rigorously in literature that only keeping the first-level images can already provide a \emph{necessary condition} to predict the occurrence of LCA. 
So we briefly sketch a proof here. Let us denote $F= F_{\mathrm{coul}} + F_1 + F_2 + \ldots$, where $F_{\mathrm{coul}}$ is the bare Coulomb force, and $F_i$ denotes the force contributed from the $i$-th level reflected images. Clearly, since $\epsilon_{1,2} > \epsilon_{\mathrm{out}}$, $F_i$ forms an alternating series, which decays to zero as the reflection level grows. By the \emph{alternating series remainder theorem}, if one truncates the summation at $F_n$, then the remainder has the same sign as the first neglected term $F_{n+1}$. Now if LCA does not occur by keeping the first-level images, which means that $F \approx F_{\mathrm{coul}} + F_1 > 0$ (repulsive) for all $R\geq a_1+a_2$, then according to the remainder theorem, the neglected polarization force $F_2 + F_3 + \ldots$ has the same sign as $F_2$, which is also repulsive, indicating that LCA will not occur even if all the higher-level image reflections are considered. This ends the proof.

To obtain a critical condition for the occurrence of LCA, we require $F=0$ at some \emph{critical separation} distance $R_c\geq a_1+a_2$ in Eq.~\eqref{force}. After simplification, we obtain the following critical condition in a dimensionless form:
\begin{equation}
	4 + k_{1}\frac{Q_{2}}{Q_{1}} H_{2}+k_{2}\frac{Q_{1}}{Q_{2}} H_{1}=0\;,
	\label{Cd}
\end{equation}
where we define two new dimensionless parameters $H_{1} = 4t_{1}+ \frac{1}{(1+t_{1})^{2}}-\frac{1}{(1-t_{1})^{2}}$ and $H_{2} = 4t_{2}+ \frac{1}{(1+t_{2})^{2}}-\frac{1}{(1-t_{2})^{2}}$ (recall that $t_{1}=\frac{a_{1}}{R_c}$, $t_{2}=\frac{a_{2}}{R_c}$). 
Eq.~\eqref{Cd} provides a general critical condition for LCA between two polarizable spheres: given a system parameter setting, where the spheres can differ in both sizes, carrying charges, and dielectric constants,
as long as Eq.~\eqref{Cd} can be satisfied for some $R_c\geq a_1+a_2$, then the theory predicts the occurrence of LCA for any $R< R_c$.
As a numerical validation, we cross-compare the prediction of $R_c$ using Eq.~\eqref{Cd} with numerical results obtained using a highly accurate hybrid method~\cite{HMgan2019efficient}. 
It is found that, over various system parameter settings, our theory will always lead to a relative error of less than $1\%$ in predicting the critical distance $R_c$, justifying the validity of our theory. 
All data comparing our theoretical prediction and benchmark numerical values are summarized in the SI, Tables S1-S2.

In what follows, we will carefully analyze the physical consequences concluded from Eq.~\eqref{Cd}. 
For the sake of clarity, we will separate our discussions into two scenarios, namely a) equal-sized spheres with both charge- and dielectric-asymmetry; and b). unequal-sized spheres with identical carrying charges and dielectric constants. 
In each scenario, we always cross-compare with numerical results to validate our theory.

\textit{Critical conditions for LCA: analysis for equal-sized spheres.}  
For equal-sized particles ($a_1=a_2=a$) with possibly different carrying charges and dielectric constants, the critical condition Eq~\eqref{Cd} reduces to,
\begin{equation}
	4 + \left(k_{1}\frac{Q_{2}}{Q_{1}}+k_{2}\frac{Q_{1}}{Q_{2}}\right)H=0\;,
	\label{Cd1}
\end{equation}
with $H = 4t+ \frac{1}{(1+t)^{2}}-\frac{1}{(1-t)^{2}}$, $t=\frac{a}{R_c}$. Clearly, due to the non-overlapping constraint, we have $t\in\left(0,\frac{1}{2}\right]$: the critical separation $R_c\to+\infty$ as $t\to0$, while they are in contact if $t=1/2$. And it can be validated that for all $t\in\left(0,\frac{1}{2}\right]$, the dimensionless parameter $H<0$.
Finally, the charge ratio $\frac{Q_{2}}{Q_{1}}$ (or $\frac{Q_{1}}{Q_{2}}$) is always positive, and $k_{1,2}=\frac{\epsilon_\text{1,2}-\epsilon_\text{out}}{\epsilon_\text{1,2}+\epsilon_\text{out}}\in(0,1)$, since here we study the interaction between like-charged polarizable spheres, thus $Q_1Q_2>0$ and $\epsilon_{1,2}>\epsilon_\text{out}>0$ always hold.
Consequently, the term $\left(k_{1}\frac{Q_{2}}{Q_{1}}+k_{2}\frac{Q_{1}}{Q_{2}}\right)$ in Eq.~\eqref{Cd1} will always be positive, and we know $H<0$, so it is possible to find a critical separation $R_c$ to make Eq.~\eqref{Cd1} hold. 
It is worth noting that if the medium is more polarizable, i.e., $\epsilon_\text{out}>\epsilon_{1,2}>0$, then $\left(k_{1}\frac{Q_{2}}{Q_{1}}+k_{2}\frac{Q_{1}}{Q_{2}}\right)<0$, and since $H<0$, which means that Eq.~\eqref{Cd1} does not hold for any $R_c$, thus the theory predicts that there is no LCA under such condition, which is consistent with previous findings\cite{MEbichoutskaia2010electrostatic,CDlindgren2016progress}.

\begin{figure}[htbp]
	\centering
	\includegraphics[scale=0.6]{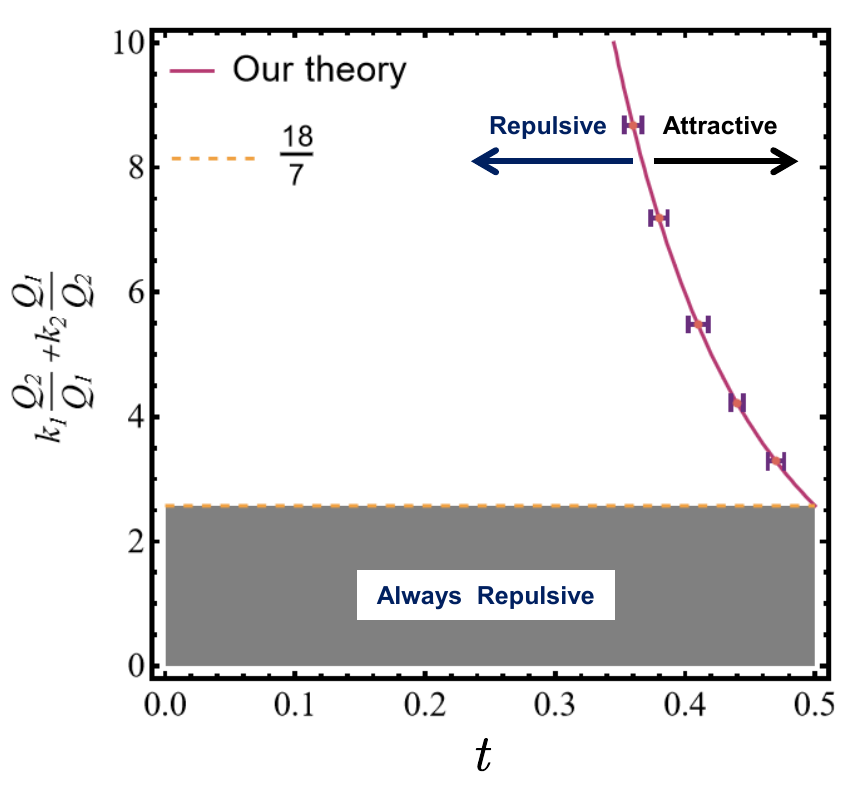}
	\caption{The critical conditions for LCA of two equal-sized spheres predicted by Eq.~\eqref{Cd1}. 
    For the case $\left(k_{1}\frac{Q_{2}}{Q_{1}}+k_{2}\frac{Q_{1}}{Q_{2}}\right)>18/7$, LCA will occur, purple curve indicates the location of $R_c$ according to Eq.~\eqref{Cd1}, while each of the error bars (in $t$) are obtained from numerical simulations of 3 different system parameter settings for the same value of $\left(k_{1}\frac{Q_{2}}{Q_{1}}+k_{2}\frac{Q_{1}}{Q_{2}}\right)$(see SI, Table S1 for detailed data). 
    The theory predicts no LCA for $\left(k_{1}\frac{Q_{2}}{Q_{1}}+k_{2}\frac{Q_{1}}{Q_{2}}\right)\leq 18/7$.}
	\label{CD1}
\end{figure}

To better illustrate the physical interpretation of Eq.~\eqref{Cd1}, we plot it by treating it as an implicit function in terms of $\left(k_{1}\frac{Q_{2}}{Q_{1}}+k_{2}\frac{Q_{1}}{Q_{2}}\right)$ and $t$. The result are documented in Fig.~\ref{CD1}. First, it is observed from Fig.~\ref{CD1} that LCA will not happen for $\left(k_{1}\frac{Q_{2}}{Q_{1}}+k_{2}\frac{Q_{1}}{Q_{2}}\right)\leq 18/7$, which can be easily justified by setting $t=1/2$ ($R_c=2a$) in Eq.~\eqref{Cd1}, we obtain
\begin{equation}
 k_{1}\frac{Q_{2}}{Q_{1}}+k_{2}\frac{Q_{1}}{Q_{2}}=\frac{18}{7}\;.
	\label{cd}
\end{equation}
Clearly, the two spheres are in contact when $t=1/2$, and if LCA does not happen at the closest distance (with strongest polarization), then it is understood that LCA will not occur at any sphere separation.
On the other hand, when $\left(k_{1}\frac{Q_{2}}{Q_{1}}+k_{2}\frac{Q_{1}}{Q_{2}}\right)>\frac{18}{7}$, LCA will occur for sphere separation within a critical distance $R_c$. In Fig.~\ref{CD1}, the LCA region corresponds to the right hand-side of the critical condition curve. Clearly, the LCA region grows as $\left(k_{1}\frac{Q_{2}}{Q_{1}}+k_{2}\frac{Q_{1}}{Q_{2}}\right)$ increases, highlighting that for equal-sized spheres, LCA is triggered by 1). charge-asymmetry; and 2). strong polarizability of the two spheres.

Next, we discuss two special situations.
a). If the two equal-sized spheres also have the same dielectric constants, namely, $k_{1}=k_{2}=k$, then $\left(k_{1}\frac{Q_{2}}{Q_{1}}+k_{2}\frac{Q_{1}}{Q_{2}}\right)$ would become $k\left(\frac{Q_{2}}{Q_{1}}+\frac{Q_{1}}{Q_{2}}\right)$. 
Clearly, in this situation, LCA can still occur, as long as the charge ratio exceeds some critical value, so that $k\left(\frac{Q_{2}}{Q_{1}}+\frac{Q_{1}}{Q_{2}}\right)>18/7$;
b). if the two equal-sized spheres also having the same carrying charges, namely, $Q_{1}=Q_{2}=Q$, then $\left(k_{1}\frac{Q_{2}}{Q_{1}}+k_{2}\frac{Q_{1}}{Q_{2}}\right)$ would become $(k_{1}+k_{2})$, which is always less than $18/7$, indicating that two equal-sized and symmetrically-charged spheres will always be repulsive, regardless of their polarizability values. 
It is worth noting that for both situations, our theoretical predictions are consistent with existing studies\cite{MEbichoutskaia2010electrostatic,CDlindgren2016progress}.

Finally, to validate our theory, we have chosen five different points on Fig.~\ref{CD1} predicted by our theory, and for each point, we validate its accuracy by choosing 3 different system parameters, but with the same value of $\left(k_{1}\frac{Q_{2}}{Q_{1}}+k_{2}\frac{Q_{1}}{Q_{2}}\right)$. Then we solve for $R_c$ numerically, yielding the error bar shown in Fig.~\ref{CD1}, demonstrating an excellent agreement between our theory and numerical simulations. Detailed data are documented in Table S1 of SI.
 
\textit{Critical conditions for LCA: analysis for unequal-sized spheres.} 
For unequal-sized spheres but with the same carrying charges and polarizability, the critical condition Eq.~\eqref{Cd} reduces to
\begin{equation}
	1 + kH=0\;,
	\label{Cd2}
\end{equation}
where we recall that $H = t_{1}+t_{2}- \frac{t_{1}}{(1-t_{1}^{2})^{2}}-\frac{t_{2}}{(1-t_{2}^{2})^{2}}$, $t_{1}=\frac{a_{1}}{R_c}$, and $t_{2}=\frac{a_{2}}{R_c}$.
Here, the non-overlapping constraint $R_c\geq a_1 + a_2$ leads us to $t_1+t_2\leq 1$. Then it can be validated that $H<0$, and since $k>0$, Eq.~\eqref{Cd2} may still predict the occurrence of LCA.

\begin{figure}[htbp]
	\centering
	\includegraphics[scale=0.65]{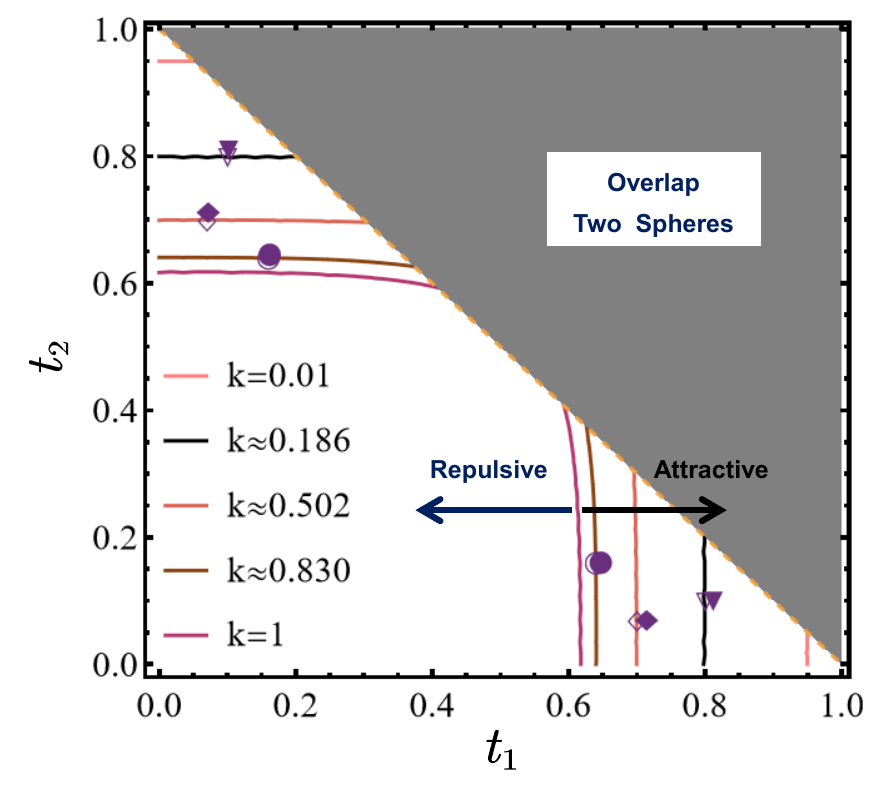}
	\caption{The critical conditions for LCA of two unequal-sized spheres predicted by Eq.~\eqref{Cd2}, under different values of $k$ ranging from $0.01$ to $1$. 
   The open and solid symbols represent theoretical and numerical results respectively, and showing an excellent agreement. 
   The upper-right shaded region has no physical meaning due to the non-overlapping constraint $t_1 + t_2\leq 1$. }
	\label{CD2}
\end{figure}

Unlike the previous case, here due to the charge-symmetry of the two spheres, LCA is expected to be triggered by 1). size-asymmetry; and 2). polarizability of the two spheres.
As a result, to better illustrate the physical interpretation of Eq.~\eqref{Cd2}, we plot it by treating it as an implicit function in terms of $t_1$ and $t_2$, and under various polarizability values of $k$ ranging from $0.01$ to $1$. The results are documented in Fig.~\ref{CD2}.
First, it is observed from Fig.~\ref{CD2} that, for different sphere polarizability $k$, there would be a critical size-ratio, characterized by $t_1/t_2$ (or $t_2/t_1$) to trigger the occurrence of LCA. 
The stronger polarizability (larger $k$) the spheres have, the less size-asymmetry is required. 
It is worth noting that, the limiting case $k=1$ corresponds to perfectly-conducting spheres ($\epsilon_{1,2}\to +\infty$), where the attraction region reaches its maximum; while the other limiting case $k\to 0$ means that $\epsilon_{1,2} \to \epsilon_{\mathrm{out}}$, where the two spheres degenerate to two point charges, in which case the attraction region also shrinks to an infinitesimal point, as can be seen in Fig.~\eqref{CD2}.

Finally, to validate our theory, we also compare our theoretical predictions with numerical simulations. 
The data points are plotted in Fig.~\eqref{CD2} as open and solid symbols, respectively, and demonstrating an excellent agreement. (Detailed data documented in Table S2 of SI). We also note that, Fig.~\ref{CD2} also validates the predicting from a previous study (Fig.4 of Ref~\cite{PIchan2020theory}), where a theory was developed to qualitatively determine the occurrence of LCA between polarizable spheres.

\textit{Conclusions and future work.} 
In summary, a novel three-point image formula is derived to calculate the interaction between two polarizable spheres. Based on this, a concise and quantitative theory is developed to predict the occurrence of LCA. Detailed analysis for the critical conditions, for both equal- and unequal-sized spheres are carried out, as well as numerical validations by cross-comparing with simulation results. The derived image formula is directly applicable to different simulation methods (MD, MC) involving the polarizable sphere model. Furthermore, the obtained critical conditions may provide physical insights into various physical and chemical processes potentially involving LCA, such as self-assembly, crystallization, and phase separation across different length scales.

In the future, we plan to extend this work to systems in which spheres are immersed in electrolyte solutions. 
In these scenarios, the ionic screening effect also becomes significant~\cite{fisher1994interaction}, and due to the additional model complexity, image charge formulas can only be obtained through approximations~\cite{deng2007discrete} or under specific limiting conditions~\cite{zhang2005long}. 
Notably, a general approach to establish semi-analytical image charge formulas
for a single dielectric sphere in electrolytes has been proposed by Z. Xu et al.~\cite{xu2013mellin}, which may offer a promising
foundation for extending our theory to the interaction between dielectric spheres inside electrolytes.
Finally, it should be mentioned that the polarization effect is of \emph{many-body} nature. Thus for the case of more than two spheres, the three-body interactions~\cite{nitzke2024long}, or more generally, many-body interactions~\cite{hassan2022manipulating,MSFfreed2014perturbative} need to be carefully investigated, which will be reserved for our future study. 

\section*{Supporting Information}\label{sec:supplementary}
Validation data for force calculations (Figures S1-S2), validation data for critical condition predictions (Tables S1-S2).

\begin{acknowledgement}
ZG would like to acknowledge financial support from the Natural Science Foundation of China (Grant No. 12201146), the Natural Science Foundation of Guangdong Province (Grant No. 2023A1515012197), the Basic and Applied Basic Research Project of Guangzhou (Grant No.2023A04J0054), and the Guangzhou-HKUST(GZ) Joint Research Project (Grant Nos. 2023A03J0003 and 2024A03J0606). 
Both authors would like to thank Prof. Ho-Kei Chan for insightful discussions about like-charge attraction.
\end{acknowledgement}

\providecommand{\latin}[1]{#1}
\makeatletter
\providecommand{\doi}
  {\begingroup\let\do\@makeother\dospecials
  \catcode`\{=1 \catcode`\}=2 \doi@aux}
\providecommand{\doi@aux}[1]{\endgroup\texttt{#1}}
\makeatother
\providecommand*\mcitethebibliography{\thebibliography}
\csname @ifundefined\endcsname{endmcitethebibliography}  {\let\endmcitethebibliography\endthebibliography}{}



\end{document}